\documentclass[10pt,letterpaper]{article}
\usepackage{opex3}
\usepackage{graphicx}
\usepackage{amsmath} 
\usepackage{amssymb}  
\usepackage[sort]{cite}  

\begin{document}

\title{High precision wavelength estimation method for integrated optics}

\author{R.~M.~Oldenbeuving,$^{1,2,3}$ H. Song,$^{4,5,6,*}$
G. Schitter,$^{7}$ M. Verhaegen,$^{5}$\\
E. J. Klein,$^{8}$ C. J. Lee,$^{1,2,9}$ H. L. Offerhaus$^{2,10}$ and
K.-J. Boller$^{1,2}$}

\address{$^1$University of Twente, Laser Physics and Nonlinear Optics Group, PO Box 217, 7500 AE, Enschede,
The Netherlands\\
$^2$MESA+ Research Institute for Nanotechnology, PO Box 217, 7500
AE, Enschede, The Netherlands\\
$^3$Satrax B.V., PO Box 456, 7500 AL, Enschede, The Netherlands\\
$^4$State Key Lab of Modern Optical Instrumentation, Zhejiang
University, 310027, Hangzhou, China\\
$^5$Delft Center for Systems and Control, Delft University of
Technology, Mekelweg 2, 2628 CD, Delft, The Netherlands\\
$^6$Ocean College, Zhejiang University, Yuhangtang Road 866, 310058, Hangzhou, China\\
$^7$Automation and Control Institute, Vienna University of
Technology, Gusshausstrasse 27-29, A-1040, Vienna, Austria\\
$^8$XiO Photonics, PO Box 1254, 7500 BG, Enschede, The
Netherlands\\
$^9$FOM Institute DIFFER, Edisonbaan 14, 3439 MN, Nieuwegein, The
Netherlands\\
$^{10}$University of Twente, Optical Sciences Group, PO Box 217,
7500 AE, Enschede, The Netherlands}

\email{$^*$hongsong@zju.edu.cn} 


\begin{abstract}
A novel and simple approach to optical wavelength measurement is
presented in this paper. The working principle is demonstrated using
a tunable waveguide micro ring resonator and single photodiode. The
initial calibration is done with a set of known wavelengths and
resonator tunings. The combined spectral sensitivity function of the
resonator and photodiode at each tuning voltage was modeled by a
neural network. For determining the unknown wavelengths, the
resonator was tuned with a set of heating voltages and the
corresponding photodiode signals were collected. The unknown
wavelength was estimated, based on the collected photodiode signals,
the calibrated neural networks, and an optimization algorithm. The
wavelength estimate method provides a high spectral precision of
about 8~pm (5$\cdot$10$^{-6}$ at 1550~nm) in the wavelength range
between 1549~nm to 1553~nm. A higher precision of 5~pm
(3$\cdot$10$^{-6}$) is achieved in the range between 1550.3~nm to
1550.8~nm, which is a factor of five improved compared to a simple
lookup of data. The importance of our approach is that it strongly
simplifies the optical system and enables optical integration. The
approach is also of general importance, because it may be applicable
to all wavelength monitoring devices which show an adjustable
wavelength response.
\end{abstract}



\section{Introduction}
Measuring and controlling the wavelength of lasers is essential to a
vast number of applications. Examples range from multichannel
wavelength division-multiplexing (WDM)~\cite{Gu1998}, optical
communication~\cite{Ophir:12}, linear and nonlinear spectroscopic
applications~\cite{Wysocki2005,Lindsay2005,Cooper1995,Richard2008},
to laser based metrology~\cite{Udem2002}. It is desirable to combine
high precision and simplicity with small size and the option of
integration.

State-of-the-art wavemeters, such as double-folded Michelson
interferometers~\cite{Banerjee2001}, can readily provide a high
spectral resolution of better than 10$^{-6}$. However, this comes at
the cost of a fairly large instrument size (in the order of 10$^6$
wavelengths, i.e., typically a meter).

There are also simplified wavemeters which are suitable for
miniaturization. Their working principle is based on two (or
several) channels in which the transmission \emph{vs}. wavelength is
different~\cite{Wang2006}. If the mapping between the transmission
of the channels and the wavelength is bijective (i.e., the
transmission curve of each channel is strictly increasing or
decreasing), the normalized transmission ratio of the channels can
be calibrated in the form of a look-up table (LUT). Miniaturization
can be achieved by using the transmission functions of integrated
optical wavelength filters~\cite{Kiesel2006}, such as thin-film
interference filters, photonic crystal
waveguides~\cite{Viasnoff2005}, or multimode interference
couplers~\cite{Sookdhis2003}. This also allows the laser and its
wavelength monitor to be integrated into a single
device~\cite{Mason1998}. However, the operational range is confined
to the region where the spectral sensitivities of all channels are
either strictly increasing or decreasing.

In this paper, we present a new method for wavelength and power
estimation, based on the calibrated transmission spectra of a
micro-ring resonator (MRR), providing high precision and extended
measurement range suitable for integrated optics. A wavemeter was
constructed in an integrated optical design for verification of the
method, which simply consists of a single, on-chip tunable MRR as an
optical transmission filter and a single photodiode. The MRR used
here is fabricated from Si$_3$N$_4$/SiO$_2$ with a box-shaped
waveguide cross section (TriPleX$^{TM}$)~\cite{Morichetti2007}. The
wavelength and power of the incident light is estimated by analyzing
the photodiode signals with neural networks and a nonlinear
optimization algorithm.

The contribution of our work lies in the method for high-precision,
range-extended wavelength and power estimation, which simplifies the
optical requirements for design, fabrication and integration of the
devices. Furthermore, with the proposed algorithm to estimate the
wavelength, the spectral sensitivity of the MRR is not necessarily
required to be strictly increasing or decreasing in the operational
range. Therefore, the operational range is extended and more tunings
can be applied for wavelength estimate (i.e., more heating voltages
can be applied to the heater of the MRR), which results in a higher
precision in the measurement. An advantage of this wavelength
estimation method is that monotonicity is not required, allowing one
to consider a larger number of physical implementations, including
devices with randomly varying transmission. Thus, one may consider
the transmission spectrum of a powder~\cite{Boer1992}, or a
multi-mode fiber~\cite{Redding2012}, for example, as long as the
transmission function is reproducible.

\section{Principle of wavelength and power estimation}
\begin{figure}[!h]
\centering
\includegraphics[width=.6\linewidth]{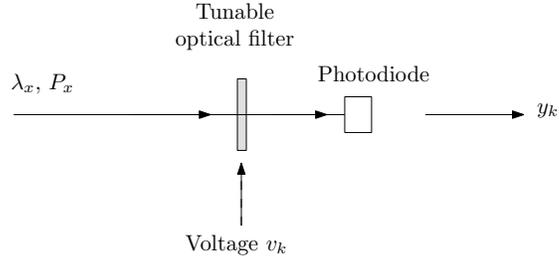}
\caption{\label{Fig:schematic} General scheme of the wavemeter.
Laser light with wavelength $\lambda_x$ and power $P_x$ passes the
tunable color filter, i.e., a micro-ring resonator (MRR) in this
paper. The spectral transmission function of the MRR can be changed
by applying a heater voltage $v_k$. The transmitted light is
collected by a photodiode, which yields a measurement signal $y_k$.
}
\end{figure}

\begin{figure}[!h]
\centering
\includegraphics[width=.7\linewidth]{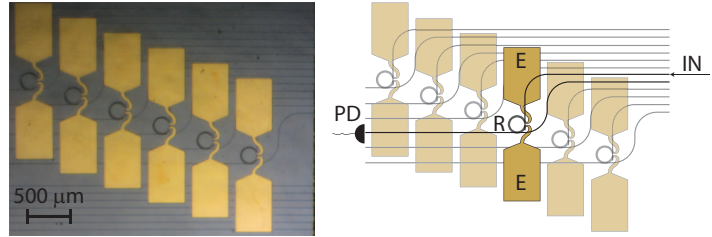}
\caption{\label{Fig:mrr_photo} Microscope picture of the waveguide
chip (left) with micro-ring resonators (MRRs) and a schematic
drawing (right) of the same chip. "R" denotes the MRR with the
heater on top (gray). "E" denotes electrical contacts (gold colored)
and the black lines represent waveguides. The fiber from the tunable
laser was connected to the MRR from "IN". "PD" denotes the position
of the photodiode. Since only one MRR was used at the same time,
there was no crosstalk between the MRR and the heaters of
neighboring MRRs.}
\end{figure}

Fig.~\ref{Fig:schematic} shows the schematic of the wavemeter. Laser
light with an unknown wavelength $\lambda_x$ and unknown power $P_x$
passes through a tunable optical filter. The transmission of the
filter \emph{vs.} wavelength, $f(\lambda,v_k)$, can be modified with
an external control parameter, $v_k$. In our case, the filter is a
single MRR equipped with an electric heater (see
Fig.\ref{Fig:mrr_photo}). The control parameter is the voltage
applied to the heater, which tunes the optical resonator length of
the MRR and thereby modifies its transmission function. To estimate
the wavelength of the input light field, a number of $N$ different
voltages are applied, where we define $v_k$ as the $k^{th}$ voltage.
The transmitted light illuminates a photodiode, which yields a set
of measurement signals $y_k$, in our case voltages. The sensitivity
of the photodiode is denoted as $d(\lambda)$, which is a function of
the wavelength $\lambda$. The measurement signal $y_k$ is given by
the product of the incident laser power, the transmission of the
MRR, and the sensitivity of the photodiode. When including
measurement noise, $\eta_k$, the measurement signal can be expressed
as
\begin{align}
y_k &= P_x f(\lambda_x,v_k)d(\lambda_x) + \eta_k =
P_xS(\lambda_x,v_k) + \eta_k. \label{eq1}
\end{align}
Here we have abbreviated the product of the MRR transmission and
photodiode sensitivity as the combined spectral sensitivity function
$S(\lambda_x,v_k)= f(\lambda_x,v_k)d(\lambda_x)$.

To obtain the unknown wavelength, $\lambda_x$, and power, $P_x$, of
incident light from measurements of $y_k$, we proceed in two steps:
\begin{enumerate}
\item Device calibration\\
The first step is to obtain information on the spectral sensitivity
function. In principle, $S(\lambda,v)$ may be derived via physical
modeling of the MRR transmission and the photodiode response.
However, in practice, the highest accuracy is achieved via
calibration. Calibration requires injecting light at a known power
at a number of known input wavelengths, $\lambda$, into the MRR. For
each wavelength, $\lambda$, the heating voltage, $v_k$, of the MRR
is tuned to $N$ values, i.e., $k=1,\cdots,N$ with $N$ the number of
tunings; and the corresponding photodiode signals, $y_k$,
($k=1,\cdots,N$) are recorded for each pair of $\lambda$ and $v_k$.

Fig.~\ref{Fig:spectracurves} shows typical sets of photodiode
measurement data. Each data set is recorded at a different heater
voltage $v_k$. An approximation for $S(\lambda,v)$ can then be
obtained by fitting an analytical function to the
data~\cite{Sjoberg:95}, such as a polynomial, a spline or a neural
network (NN). A 2-layer neural network is able to model a broad
range of nonlinearities~\cite{Lippmann1987,Kung:93,Haykin:94} and,
from a practical point of view, can be implemented and trained very
conveniently, e.g., with a neural network toolbox~\cite{MatlabNN07}.
Therefore a 2-layer neural network with $Q$ neurons in the first
layer and one in the second layer was chosen for our work. At a
certain control voltage $v_k$, the output $y_k(\lambda)$ of the
neural network is given as

\begin{equation}\label{eqNN}
y_k(\lambda)=P \cdot
\hat{S}(\lambda,v_k)=P\displaystyle\sum\limits_{i=1}^Q
w_{1ik}tanh(w_{2ik}\lambda+s_{1ik})+s_{2ik}.
\end{equation}

Eq.~(\ref{eqNN}) models the spectral sensitivity curve as a
superposition of step response functions (neurons), in this case
chosen as tangent hyperbolic functions. $w_{1ik}$ and $w_{2ik}$ are
the input and output weights of the neural network, respectively;
$s_{1ik}$ and $s_{2ik}$ are biases on the input and output neurons,
respectively. To obtain a good compromise between a best fit and a
minimum number of fit parameters, the number of neurons, $Q$, should
be selected according to the required fitting accuracy of the neural
network~\cite{Haykin:94,MatlabNN07}.

\begin{figure}
\centering
  \includegraphics[width=.6\linewidth]{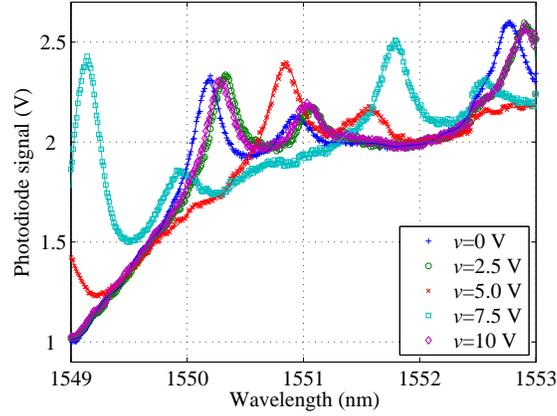}\\
     \caption{Spectral sensitivity function $S(\lambda,v)$ at different $v$. "+": calibration data; lines: curves fitted by neural networks.
     For clarity, curves are shown only for five heating voltages $v=$0~V, 2.5~V, 5.0~V, 7.5~V and 10~V.}
     \label{Fig:spectracurves}
\end{figure}

\item Wavelength and power estimation\\
In the second step, light with an unknown wavelength, $\lambda_x$,
and power, $P_x$, is sent into to the wavemeter. A set of \emph{N}
photodiode signals, $y_k$, is recorded via applying \emph{N}
different heater voltages, $v_k$ ($k=1,\cdots,N$). The heater
voltages can be equally spaced (as in the experiments described in
this paper), randomly selected or designed by the user beforehand.
For each heater voltage, $v_k$, a nonlinear equation, as described
in Eq.~(\ref{eq1}) can be formed. By stacking $N$ equations, a set
of nonlinear equations is obtained as
\begin{align}\label{eqNLset}
Y \equiv  \left[\begin{array}{ccc}
                                            y_1\\
                                          y_2\\
                                          \vdots\\
                                          y_N
                                          \end{array}\right] =
                         \left[ \begin{array}{ccc}
                                            P_xS(\lambda_x,v_1)+ \eta_1 \\
                                          P_xS(\lambda_x,v_2)+ \eta_2 \\
                                          \vdots\\
                                          P_xS(\lambda_x,v_N)+ \eta_N \\
                                          \end{array}
                           \right] .
\end{align}
The unknown wavelength, $\lambda_x$, and power, $P_x$, can be
estimated by solving Eq. (\ref{eqNLset}). Obtaining an analytical
solution of Eq.(\ref{eqNLset}) may be infeasible in practice.
Alternatively, a numerical solution can be obtained by solving a
nonlinear least squares (NLLS) problem as
\begin{align}\label{eqfsolve}
\left(\hat{\lambda}_x,\hat{P}_x\right) = \mathrm{arg}
\min_{\hat{\lambda}_x,\hat{P}_x} \underbrace{\frac{1}{N}\left|
Y-\hat{Y} \right|^2}_{J},
\end{align}
with $Y$ defined in Eq. (\ref{eqNLset}) and $\hat{Y}$ defined as
\begin{align}\label{eqNLsetEST}
\hat{Y} \equiv  \left[\begin{array}{ccc}
                                            \hat{y}_1\\
                                          \hat{y}_2\\
                                          \vdots\\
                                          \hat{y}_N
                                          \end{array}\right] =
                         \left[ \begin{array}{ccc}
                                            \hat{P}_x\hat{S}(\hat{\lambda}_x,v_1)\\
                                          \hat{P}_x\hat{S}(\hat{\lambda}_x,v_2) \\
                                          \vdots\\
                                          \hat{P}_x\hat{S}(\hat{\lambda}_x,v_N)\\
                                          \end{array}
                           \right] .
\end{align}
Here $\hat{S}(\lambda,v_k)$ is the approximated spectral sensitivity
function that was obtained by calibration in Step 1. The
mean-square-error between the measurement vector $Y$ and the
approximation $\hat{Y}$ at certain guesses, $\hat{\lambda}_x$ and
$\hat{P}_x$, is defined as the cost function
$J=\frac{1}{N}|Y-\hat{Y}|^2$. Eq.~(\ref{eqfsolve}) aims to find an
appropriate $\hat{\lambda}_x$ and $\hat{P}_x$ such that the
approximation $\hat{Y}$ is as close as possible to the measurement
$Y$, i.e., $J$ is minimized. In this way, the unknown wavelength,
$\lambda_x$ and power $P_x$, of the incident light field are
estimated, i.e., the wavelength readout is obtained. By accurately
approximating the spectral sensitivity, $S(\lambda,v_k)$, and
increasing the number of tunings (i.e., more equations are used),
the accuracy of $\hat{\lambda}_x$ and $\hat{P}_x$ can be improved.

\end{enumerate}

The special feature of our method is that the spectral sensitivity
function, $\hat{S}(\lambda,v_k)$, is not necessarily bijective
(i.e., strictly increasing or decreasing) with respect to
wavelength, $\lambda$, in the operational range, as compared with
the LUT method. Because of this, not only is the operational range
of our wavemeter extended (non-monotonous regions are also allowed),
but also the accuracy of the estimates is improved. Since
monotonicity is not required, sensitivity curves with local peaks
and troughs (and even multi-peaked and multi-troughed sensitivity
functions) can be used for wavelength estimate. As more equations
are added in Eq.~(\ref{eqNLsetEST}), the accuracy of the estimate
improves.

\section{Experiments and results}
Although our wavelength estimation method is not limited to one
specific tunable optical filter, for an experimental demonstration,
we used an available MRR. It consists of single-mode waveguide with
a designed waveguide width of 450~nm. The waveguide chip is covered
with a top cladding of 12~$\mu$m, upon which the heaters and
electrodes are deposited. The cross section of the waveguide is
box-shaped. The waveguide fabrication process is described in
\cite{Morichetti2007}. The radius of the MRR is R=85.47~$\mu$m and
the group index is $n_g=$1.73. The coupling gap between the straight
waveguide and the MRR is about 1.1~$\mu$m, resulting in a power
coupling coefficient of about 0.26. This corresponds to a free
spectral range of about 2.6~nm and a $Q$-factor of about 6000. The
through port of the MRR is connected to an independently calibrated,
fiber-coupled tunable laser source (HP81689A, Agilent). The light
transmitted through the MRR is coupled directly to a photodiode
(FGA10, Thorlabs). The signal of the photodiode (i.e., $y_k$) is
amplified and recorded by a computer via a data acquisition card
(PCIe-6251, National Instruments). The heater voltage $v_k$ is
provided by the computer and a power amplifier.

To calibrate the sensitivity function, $S(\lambda,v)$, and to test
the wavelength estimate method, two data sets were collected,
denoted as $\{\lambda,v,y_c\}$ and $\{\lambda,v,y_t\}$,
respectively. These two data sets were generated using a set of
known wavelengths and heater voltages. The laser was tuned over the
wavelength range from 1549~nm to 1553~nm, in 400 equal steps of
0.01~nm. The power was constant at 2~mW. For each wavelength, $N=21$
heater voltages were applied, from 0~V to 10~V in steps of 0.5~V,
and the corresponding photodiode signals were collected twice, one
for calibration (i.e., $y_c$) and the other for test (i.e., $y_t$).
Thus, in total, $401 \times 21=8421$ data points were recorded for
calibration and the other 8421 data points for testing.

The settling time of the photodiode signal is about 0.5 milliseconds
in case of a step heating voltage input. During data collection, the
original sampling rate of the data acquisition card is 1~kHz. To
reduce the noise in the photodiode signal, 100 original samples are
averaged to generate one data point. Readout takes about 0.04
seconds. Therefore getting one data point takes about 0.14 seconds.
Because the communication between the control computer and the
tunable laser and the tuning of the laser wavelength both take
seconds, a typical time required for collecting both the calibration
and test data sets is about two hours.

It can be seen in Fig.~\ref{Fig:spectracurves}, that
$S(\lambda,v_k)$ consists of a broad range of nonlinear variations
\emph{vs.} wavelength, resembling a set of mutually shifted
Fabry-Perot resonator transmission functions. At longer wavelengths,
the depth of the modulation decreases. This can be attributed to the
overall heating of the optical waveguide chip, which causes the
alignment between the optical filter and the waveguide to change
over the course of the measurements.

\begin{figure}[!h]
\centering
  \includegraphics[width=.6\linewidth]{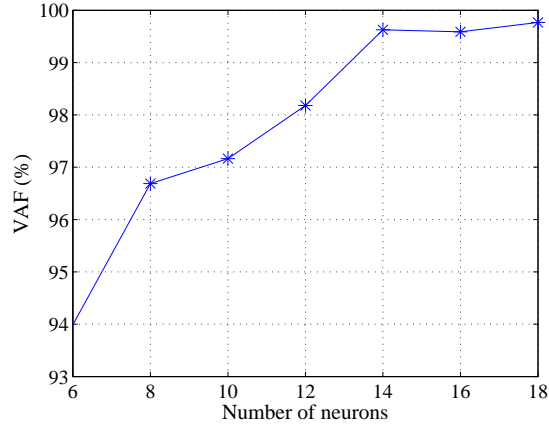}\\
     \caption{Variance accounted for (VAF) of the neural networks with respect to the number of neurons ($Q$).
     The VAF in the vertical axis is averaged over VAF of all 21 neural networks.
     As more neurons are used, the accuracy of the neural network increases until $Q=14$.}
     \label{Fig:VAFwrtQ}
\end{figure}

During calibration, each sensitivity function at a particular
voltage as a function of wavelength is approximated by a neural
network, i.e., 21 neuronal network fits are performed, one for each
heater voltage. To select an appropriate number of neurons, $Q$, in
the network as presented by Eq. (\ref{eqNN}) (2 layers, with $Q$
neurons in the first layer and one neuron in the second layer), the
\emph{variance accounted for} (VAF) was used as a
criterion~\cite{Verhaegen07}. The VAF is defined by
\begin{align}\label{eqVAF}
    A(\hat{y}_k,{y}_k)=\left(1-\frac{\mathrm{var}(\hat{y}_k-{y}_k)}{\mathrm{var}({y}_k)}\right)\times 100\%.
\end{align}
Here, $A(\hat{y}_k,{y}_k)$ is the VAF between $\hat{y}_k$ and $y_k$,
ranging from $-\infty$ to 100\%. var($y_k$) is the variance of
$y_k$. Fig.~\ref{Fig:VAFwrtQ} shows the VAF of neural networks with
respect to $Q$. As more neurons are used, the accuracy of the neural
network increases until $Q=14$. For $Q>14$, the fitting accuracy is
limited by the measurement noise in the data. Accordingly we decided
to use 14 neurons in our network. The corresponding VAF (averaged
over all 21 neural networks) is 99.63\%, indicating an accurate
approximation of the spectral sensitivity functions. This can be
seen in Fig.~\ref{Fig:spectracurves}, which shows an excellent
agreement between the neural network fitting functions and the
experimental data.

After calibration, the wavelength and power estimate algorithm was
tested with the data set $\{\lambda,v,y_t\}$, where $\lambda$ is
considered unknown. For each "unknown" wavelength $\lambda_{x,i}$,
$i\in [1,401]$, the corresponding photodiode signals $y_{t,i,k},
k=1,\cdots,21$ are scaled by a random number $p_{x,i}\in [0.1, 10]$.
This simulates an input that has both an unknown wavelength and an
unknown power, which are to be estimated. The set of scaled
photodiode signals are given as $\tilde{y}_{t,i,k}=p_{x,i}y_{t,i,k},
k=1,\cdots,21$. Heating voltages, $v_k$, and signals,
$\tilde{y}_{t,i,k}$, were then fed into the wavelength estimation
algorithm and the unknown wavelength $\lambda_{x,i}$, and
power-scaling factor $p_{x,i}$ were estimated.

\begin{figure}[!tb]
\centering
  \includegraphics[width=.6\linewidth]{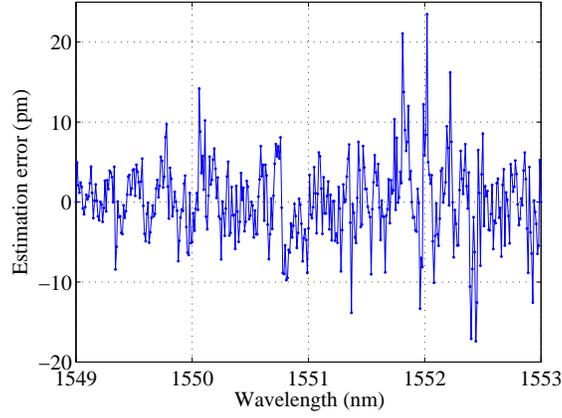}\\
     \caption{Wavelength estimation error $\hat{\lambda}_x-\lambda_x$ in the test set.}
     \label{fgErrorVALSET1}
\end{figure}

\begin{figure}[!tb]
\centering
    \includegraphics[width=.6\linewidth]{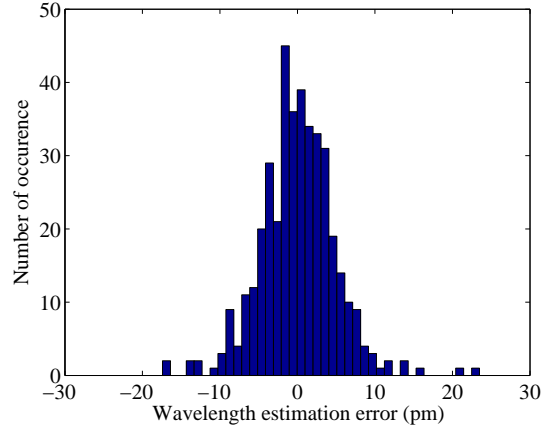}\\
     \caption{Histogram of the wavelength estimation error. The maximal estimation error for the test set is 22~pm. 95\% of
the wavelength estimation errors are below 8~pm in the range between
1549~nm and 1553~nm.}
     \label{fgErrorVALSET2}
\end{figure}

\begin{figure}[!tb]
\centering
        \includegraphics[width=.6\linewidth]{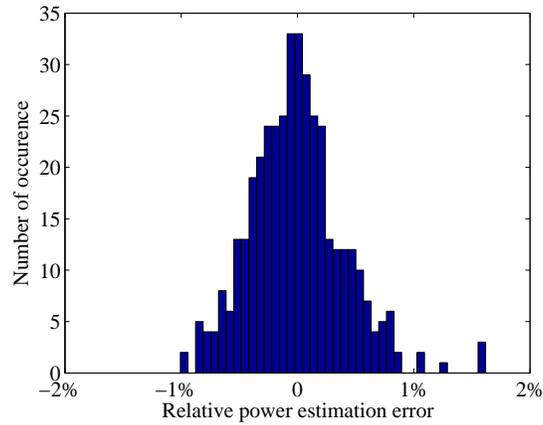}\\
     \caption{Histogram of the relative power estimation error $(\hat{p}_{x}-p_{x})/p_x$ in
     the test data set. The relative power estimation error is at maximum 1.6\%, and in 95\% of the cases less than 0.8\%.}
     \label{fgErrorPower}
\end{figure}

Fig.~\ref{fgErrorVALSET1} shows the wavelength estimation error
$\hat{\lambda}_x-\lambda_x$ with respect to wavelength, $\lambda_x$,
and Fig.~\ref{fgErrorVALSET2} shows the estimation error in a
histogram of occurrence. It can be seen that the maximum absolute
wavelength estimation error for the test set is 22~pm and that 95\%
of the wavelength estimation errors are below 8~pm in the range from
1549~nm to 1553~nm. Fig.~\ref{fgErrorPower} shows the histogram of
the relative power estimation error $(\hat{p}_{x}-p_{x})/p_x$ with
$\hat{p}_{x}$ the estimate of $p_x$ . The error in power estimation
is 1.6\% at maximum with respect to the known value; and in 95\%
cases, the power estimation error is less than 0.8\% with respect to
the known value.

\begin{figure}[!tb]
\centering
  \includegraphics[width=.6\linewidth]{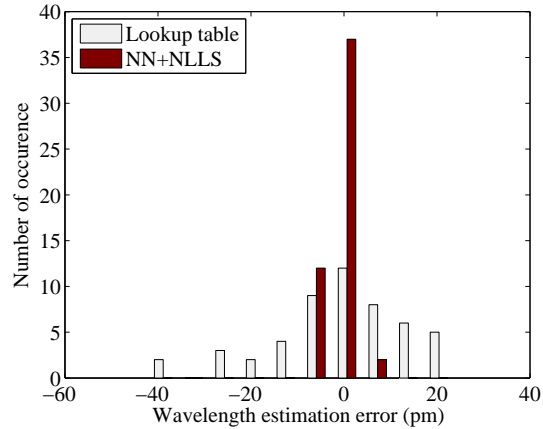}\\
     \caption{Comparison between LUT and the proposed NN+NLLS method.
95\% of the errors fall below 27~pm for LUT and 5~pm for the
proposed method (about one fifth of that for LUT).}
     \label{fgCompare}
\end{figure}

To analyze whether the proposed neural network and nonlinear least
square (NN+NLLS) method provides improved accuracy over a simple LUT
method, we chose a LUT where the ratio between two spectral
sensitivity curves strictly decreases. For the MRR used here, we
found a spectral region that satisfied this constraint, e.g.,
between 1550.3~nm and 1550.8~nm for the heater voltages $v_k$=0~V
and 5~V. The wavelength was estimated using the LUT method and for
the same data using the NN+NLLS method, as described above. The
methods are compared in Fig.~\ref{fgCompare}. The maximum wavelength
estimation error in the named range is 42~pm for the LUT and 6~pm
for our NN+NLLS approach. 95\% of the errors are less than 27~pm for
the LUT and less than 5~pm for NN+NLLS. This shows the NN+NLLS
method, indeed, provides improved accuracy over that of LUT. The
latter corresponds to a relative precision of about
3$\cdot$10$^{-6}$. The improvement is as big as a factor of five.

\section{Discussion}
The accuracy of the wavelength and power estimation depends on not
only on the accuracy of the neural networks, but also on the number
of data points used for estimation and the noise in the data.

\begin{figure}[!h]
\centering
  \includegraphics[width=.6\linewidth]{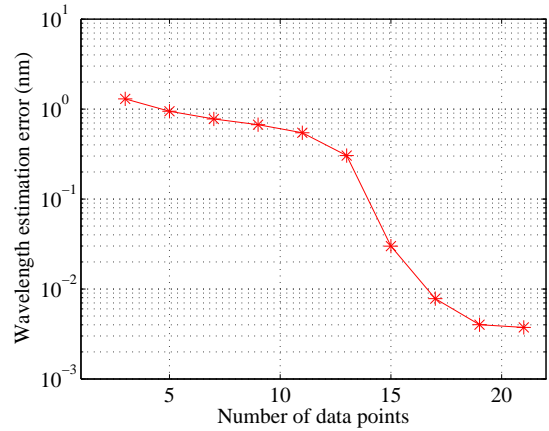}\\
     \caption{Mean of the wavelength estimation error
     decreases as more data points are used in the estimation.}
     \label{fgErrwrtN}
\end{figure}

As more data points are used, more equations can be added to the
nonlinear equation set and the accuracy of wavelength and power
estimation will be improved. To verify this, wavelength and power
estimation has been carried out where the number of data points used
increases from 3 to 21 at a step of 2 (i.e., $N$=3, 5, $\cdots$,
21). The result is shown in Fig.~\ref{fgErrwrtN}. It can be seen
that the mean of the wavelength estimation error decreases as more
data points are used as expected. There is a significant drop in the
estimation error when more than 15 data points are used and the mean
of the wavelength estimation error is limited to around 4~pm, which
is mainly due to the modeling error of the neural networks and the
noise in the measurement.

\begin{figure}[!h]
\centering
  \includegraphics[width=.6\linewidth]{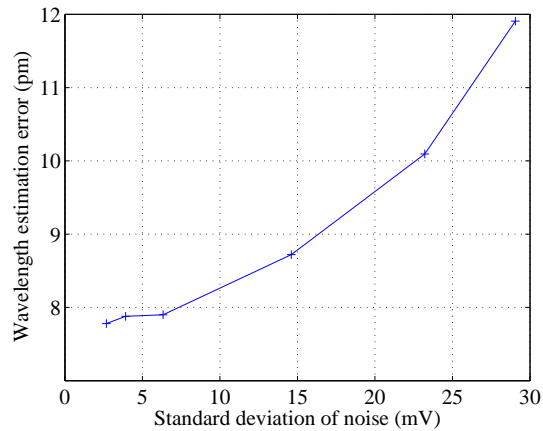}\\
     \caption{Influence of noise on the wavelength estimation error. 95\% of
the wavelength estimation error is still within 10~pm when the
standard deviation of the noise is 22.5~mV}
     \label{fgnoisetol}
\end{figure}

Currently, the noise between the calibration and test data points
has a mean value of 0.054~mV and standard deviation of 2.7~mV. To
show how our algorithm is sensitive to the noise, a simulation has
been carried out, where random noise has been added to the test data
artificially and the wavelength and power are estimated. Simulation
result has been shown in Fig.~\ref{fgnoisetol}, which indicates that
95\% of the wavelength estimation error is still within 10~pm when
the standard deviation of the noise is 22.5~mV.



\section{Conclusions}
In conclusion, we have presented a novel method to measure the
unknown power and wavelength of a laser with high precision and an
extended measurement range. The method has been verified in a simple
setup, based on an integrated optics micro ring resonator, which
allows the wavemeter to be integrated with other on-chip optical
components. Neural networks were used to approximate the spectral
sensitivity functions of the optical channel at different heating
voltages. When injecting light with an unknown wavelength and power,
nonlinear equations are formed and solved to provide an estimate of
the wavelength and power. In our experimental verification, we
demonstrated a spectral precision of about 8~pm (5$\cdot$10$^{-6}$)
in the wavelength range between 1549~nm and 1553~nm. A higher
precision of 5~pm (3$\cdot$10$^{-6}$) was achieved in the range
between 1550.3~nm and 1550.8~nm, which is a factor of five compared
to a simple lookup of data.

The advantage of our approach is that it does not attempt to obtain
a high precision and an increased wavelength range with a
sophisticated and highly precise design of an appropriate spectral
response function. Instead, our approach is able to handle all
deviations that may enter via fabrication processes or fiber input
and output coupling, as long as the optical transmission function is
reproducible. Essentially, the approach replaces the need for a
complex and precise optical design with a smart readout ("trading
hardware for software"). This can help to reduce the cost of
fabrication and expand the range of applications.

\section{Acknowledgement}
This work is supported by the STW project Smart Optical Systems:
Waveguide Based External Cavity Semiconductor Laser Arrays (project
NO. 10442) and in part by the Program for Zhejiang Leading Team of
S\&T Innovation (project NO. 2010R50036).

\end{document}